\documentclass[aps,prr,twocolumn,groupedaddress,amsmath,amssymb,amsfonts,floatfix]{revtex4-2}
\usepackage{txfonts}
\usepackage{bm} 
\usepackage{subfigure}
\usepackage{mathrsfs}
\usepackage{array}
\usepackage{amsmath,latexsym}
\usepackage{amsfonts} 
\usepackage{amssymb} 
\usepackage{braket}
\usepackage{comment} 
\usepackage{graphicx, color}
\usepackage[colorlinks=true, bookmarks=true,
bookmarksnumbered=true, bookmarkstype=toc, linkcolor=magenta,
urlcolor=cyan, citecolor=cyan]{hyperref}
\begin{document}
\title{Teacher-student learning for a binary perceptron with quantum fluctuations}
\author{Shunta Arai$^{1,2}$}
\email[]{shunta.arai.s6@dc.tohoku.ac.jp}
\author{Masayuki Ohzeki$^{1,2,3}$}
\author{Kazuyuki Tanaka$^1$}
\affiliation{$^1$Graduate School of Information Sciences, Tohoku University, Sendai 980-8579, Japan\\
$^2$Sigma-i Co., Ltd., Tokyo, Japan\\
$^3$Institute of Innovative Research, Tokyo Institute of Technology, Nagatsuta-cho 4259, Midori-ku, Yokohama Kanagawa, 226-8503 Japan}
\date{\today}

\begin{abstract}
We analyse the generalisation performance of a binary perceptron with quantum fluctuations using the replica method.
An exponential number of local minima dominate the energy landscape of the binary perceptron.
Local search algorithms often fail to identify the ground state of the binary perceptron.
In this study, we consider the teacher-student learning and
compute the generalisation error of the binary perceptron with quantum fluctuations.
Due to quantum fluctuations, we can efficiently find robust solutions that have better generalisation performance than the classical model. 
We validate our theoretical results through quantum Monte Carlo simulations.
We adopt the replica symmetry (RS) ansatz  and static approximation. 
The RS solutions are consistent with our numerical results, except for the relatively low strength of the transverse field and high pattern ratio. 
These deviations are caused by the violation of ergodicity and static approximation. 
After accounting for the deviation between the RS solutions and numerical results, the enhancement of generalisation performance with quantum fluctuations holds.
\end{abstract}
\maketitle
\section{\label{sec:sec1}Introduction}
Deep neural networks (DNNs) have achieved excellent performance on a wide range of tasks \cite{Goodfellow_book}. 
In supervised learning, DNNs can generalise for unseen data better than classical machine learning algorithms. 
Generally, when applying DNNs for supervised learning, the number of parameters is greater than the number of datapoints, which is called an `over-parameterised setting'.  
In such a setting, it is difficult to explain why DNNs generalise well using classical statistical learning theory \cite{Belkin2019}.
The theoretical understanding of the generalisation performance of DNNs is under development \cite{Bahri_2020}.

For simple models, typical generalisation performance can be theoretically analysed using statistical mechanics \cite{engel_book,nishimori_book}. 
The simplest case is the binary perceptron \cite{Gardner_1988a,Gardner_1988b}. 
In statistical mechanics analysis, we consider the thermodynamic limit as the number of parameters $N\rightarrow \infty$ and number of data $p\rightarrow \infty$ while fixing the number of data per parameter $\alpha\equiv p/N=O(1)$. 
For random pattern learning, a binary perceptron can learn random input-output patterns until reaching the storage capacity of $\alpha_c\simeq 0.833$ \cite{Krauth_1989}.
For teacher-student learning, we consider a student perceptron and teacher perceptron.
The teacher perceptron generates output data from input data and the student perceptron learn input-output patterns for a given dataset.
The student perceptron can predict the outputs generated by the teacher perceptron with $\alpha\geq \alpha_c\simeq 1.245$ \cite{Gyogyi_1990}.
The energy landscape of the binary perceptron is highly non-convex and is dominated by an exponential number of local minima \cite{Huang_2013, Huang_2014}.
Ground states are geometrically isolated in the solution space.
The hamming distance between each solution is
proportional to $N$. 
A local spin-flip algorithm based on free energy minimisation such as simulated annealing (SA) \cite{Kirkpatrick_1983} can easily become stuck in metastable states and fail to find the ground state \cite{Horner_1992, Patel_1993}.
In the solution space, there is a subdominant dense region in which the solutions have high local entropy.
In this region, all solutions have similar energy values.
We refer to these solutions as robust solutions in this paper. 
By incorporating local entropy, we can modify the standard SA technique.
The resulting algorithm can find subdominant solutions \cite{Baldassi_2016a, Baldassi_2016b}.
The quality of subdominant solutions is different from that of the typical solutions found by standard SA. 
In teacher-student learning, the generalisation error of the subdominant solutions is lower than that of typical solutions \cite{Baldassi_2015}.
 
Several empirical studies have demonstrated that the local energy (loss) landscape is related to the generalisation performance \cite{keskar_2017, Laurent_2017,pittorino2020}.
Stochastic gradient descent (SGD) with small batches often finds a flat minimiser within the energy landscape.  
In contrast, SGD with large batches converges to a sharp minimiser.
Although the training errors of these two solutions are the same or similar, the generalisation performances differ significantly. 
A flat minimiser generalises better than a sharp minimiser.
A flat region in the energy landscape is robust to perturbations in parameters and data fluctuations.
By incorporating the effects of entropy into a loss function,
a flat minimiser is algorithmically reachable.
The Entropy-SGD algorithm developed in a recent study \cite{Chaudhari_2019} computes local entropy and exploits stochastic gradient Langevin dynamics \cite{welling_2011} to approximate the gradient of local entropy, 
allowing it to find a flat minimiser.

To optimise DNNs, we can utilise quantum fluctuations.
In a recent study \cite{ohzeki2018}, the authors have formulated an optimisation algorithm based on quantum fluctuations using a path-integral representation. They have determined that finite-value quantum fluctuations improve the generalisation performance of DNNs, allowing solutions to converge to a flat minimiser. 
In the binary perceptron for random pattern learning,
quantum annealing (QA) \cite{Kadowaki_1998,Santoro_2002,Das_2008} can identify a flat region in the energy landscape \cite{Baldassi_2017}.
The local entropy obtained by QA is greater than that of typical solutions obtained by SA. 

In this study, we extend previous research on random pattern learning \cite{Baldassi_2017} to teacher-student learning.
We compute the typical behaviours of the generalisation error of the binary perceptron with quantum fluctuations by using the replica method.
In the previous study, it has been observed that quantum fluctuations lead to a flat minimiser.
In this study, we investigate whether the generalisation performance of the binary perceptron is enhanced by quantum fluctuations. 
This study is an analytical demonstration of the effectiveness of quantum fluctuations for machine learning problems.

The remainder of this paper is organised as follows.
In Section \ref{sec:sec2}, we present the formulation of the binary perceptron with quantum fluctuations. 
In Section \ref{sec:sec3}, we derive the free energy and the saddle-point equations.
In Section \ref{sec:sec4}, we numerically solve the saddle-point equations and present the phase diagrams.
To verify our theoretical analysis, we perform quantum Monte Carlo simulations.
We verify the robustness of the solutions obtained by the quantum Monte Carlo simulations using the energy landscape around the solutions.
Finally, in Section \ref{sec:sec5}, we conclude our study and discuss future research directions. 

\section{\label{sec:sec2}Binary perceptron with quantum fluctuations}
We consider teacher-student learning in a single-layer binary perceptron.
The student perceptron learns input-output patterns from the given dataset $\mathcal{D}=\{(\bm{x}_\mu,y_\mu)\}_{\mu=1}^p$.
The number of data points is represented by $p$.
For each sample, an input data vector $\bm{x}_{\mu}\in \{\pm1\}^N$ is generated from the uniform distribution $P(\bm{x}_{\mu})=\prod_{i=1}^N\left(\delta(x_{i\mu}-1)+\delta(x_{i\mu}+1)\right)/2^N$, where $N$ is the dimensionality of the input data.
The joint probability distribution of the input data is denoted as $P(\bm{X})=\prod_{\mu=1}^pP(\bm{x}_{\mu})$.
The output data are determined by the teacher perceptron as $y_\mu=\mathrm{sgn}\left(1/\sqrt{N}\sum_{i=1}^Nw_{i}x_{i\mu}\right)\in\{\pm1\}$, where $\mathrm{sgn}(\cdot)$ is the signum function. 
The weight vector of the teacher perceptron is generated from the uniform distribution as 
$P(\bm{w})=\prod_{i=1}^N\left(\delta(w_{i}-1)+\delta(w_{i}+1)\right)/2^N$.
The joint distribution of the dataset is given by 
\begin{align}
P(\mathcal{D}|\bm{w})&=\prod_{\mu=1}^pP(y_\mu|\bm{x}_\mu,\bm{w})P(\bm{x}_\mu)\nonumber\\
&=\prod_{\mu=1}^p\delta\left(y_\mu-\mathrm{sgn}\left(\frac{1}{\sqrt{N}}\sum_{i=1}^Nw_{i}x_{i\mu}\right)\right)P(\bm{x}_{\mu}).
\label{eq1}
\end{align}

The learning problem involves finding a weight vector $\bm{\sigma}=(\sigma_1,\dots,\sigma_N)\in\{\pm1\}^N$ such that all input data are simultaneously classified correctly. 
We formulate this problem as a Bayesian inference problem.
By using the Bayes formula, the posterior distribution is expressed as 
\begin{align}
P(\bm{\sigma}|\mathcal{D})&=\frac{P(\mathcal{D}|\bm{\sigma})P(\bm{\sigma})}{\sum_{\bm{\sigma}}P(\mathcal{D}|\bm{\sigma})P(\bm{\sigma})}.
\label{eq2}
\end{align}
We define the likelihood $P(\mathcal{D}|\bm{\sigma})$ as 
\begin{align}
P(\mathcal{D}|\bm{\sigma})&\propto\exp\left(-\beta E(\bm{\sigma}) \right),\label{eq3}\\
E(\bm{\sigma})&=\sum_{\mu=1}^{p}\Theta \left(-y_\mu \mathrm{sgn}\left(\frac{1}{\sqrt{N}}\sum_{i=1}^Nx_{i\mu}\sigma_i\right)\right),
\label{eq4}
\end{align}
where $\beta=1/T$ is the inverse temperature, $\Theta(x)$ is the Heaviside step function, and $\Theta(x)=1\hspace{3pt}\mathrm{if} \hspace{3pt}x>0 or \Theta(x)=0$, otherwise. The Hamiltonian $E(\bm{\sigma})$ represents the number of misclassifications.
We set the prior distribution $P(\bm{\sigma})$ to the uniform distribution $P(\bm{\sigma})=\prod_{i=1}^N\left(\delta(\sigma_{i}-1)+\delta(\sigma_{i}+1)\right)/2^N$. 
In this case, we can omit $P(\bm{\sigma})$ from Eq.\eqref{eq2}.

We often utilise the maximum a posteriori (MAP) estimation to estimate $\bm{\sigma}$.
In the MAP estimation, we find the state that maximises the posterior Eq.\eqref{eq2} at the limit of $\beta \rightarrow \infty$.
This corresponds to searching for the ground state of the Hamiltonian Eq. \eqref{eq4}.
To find the ground state, we typically adopt SA. 
However, SA can fail to identify the ground state based on the existence of many local minima.
Therefore, instead of maximising the posterior, we consider finite-temperature estimation.
At low temperatures, the probability measure in Eq.\eqref{eq2} concentrates on low-energy states. 
The learning strategy involves sampling low-energy states from Eq.\eqref{eq2} at low temperatures. 
The estimated weight vector is expressed by the expectation over the posterior distribution as $\langle\sigma_i\rangle=\sum_{\bm{\sigma}}\sigma_iP(\bm{\sigma}|\mathcal{D})$. 
 
The indicators of learning outcomes are the training and generalisation errors.
The training error is given by 
\begin{align}
\epsilon_t(\mathcal{D})&=\frac{1}{p}\langle E(\bm{\sigma}) \rangle.
\label{eq5}
\end{align} 
To evaluate performance for unseen data, we consider the generalisation error as 
\begin{align}
\epsilon_g(\mathcal{D})&=\mathbb{E}_{\{\bm{x}^{\mathrm{new}},y^{\mathrm{new}}\}}\left[\Theta\left(-y^{\mathrm{new}}\mathrm{sgn}\left(\frac{1}{\sqrt{N}}\sum_{i=1}^Nx_{i}^{\mathrm{new}}\langle\sigma_i\rangle\right)\right)\right],
\label{eq6}
\end{align}
where $\mathbb{E}_{\{\bm{x}^{\mathrm{new}},y^{\mathrm{new}}\}}[\cdot]$ denotes the expectation over  $P(\mathcal{D}^{\mathrm{new}}|\bm{w})=P(\bm{x}^{\mathrm{new}},y^{\mathrm{new}}|\bm{w})=P(y^{\mathrm{new}}|\bm{x}^{\mathrm{new}},\bm{w})P(\bm{x}^{\mathrm{new}})$. These quantities are expected to exhibit a `self-averaging' property at the thermodynamic limit $N\rightarrow\infty$. 
The observables for a quenched realisation of $\mathcal{D}$ and $\bm{w}$ are equivalent to the self-expectation over the data distribution $P(\mathcal{D}|\bm{w})P(\bm{w})$. 
For example, the generalisation error can be expressed as
\begin{align}
\lim_{N\rightarrow\infty}\epsilon_g&=[\epsilon_g(\mathcal{D})]_\mathcal{D}\nonumber\\
&=\frac{1}{\pi}\cos^{-1}m,
\label{eq7}
\end{align}
where the bracket $[\cdot]_{\mathcal{D}}$ indicates the expectation over the data distribution $P(\mathcal{D}|\bm{w})P(\bm{w})$ and $m=1/N\sum_{i=1}^Nw_i\langle\sigma_i\rangle$ denotes the overlap between the teacher and student \cite{engel_book}.

We can extend the above formulation into a quantum system as follows: 
\begin{align}
P(\mathcal{D}|\bm{\hat{\sigma}^z})&\propto\exp\left(-\beta \hat{\mathcal{H}} \right),\label{eq8}\\
\hat{\mathcal{H}}&=E(\bm{\hat{\sigma}^z})-\Gamma\sum_{i=1}^N\hat{\sigma}_i^x,\label{eq9}\\
E(\bm{\hat{\sigma}^z})&=\sum_{\mu=1}^{p }\Theta\left(-y_{\mu}\mathrm{sgn}\left(\frac{1}{\sqrt{N}}\sum_{i=1}^Nx_{i\mu}\hat{\sigma}_i^z\right)\right),
\label{eq10}
\end{align}
where $\hat{\sigma}_i^z$ and $\hat{\sigma}_i^x$ are the $z$ and $x$ components of the Pauli matrices at site $i$, respectively, and $\Gamma$ is the strength of the transverse field.
The learning strategy involves sampling low-energy states from the density matrix 
$\hat{\rho}\equiv e^{-\beta \hat{\mathcal{H}}}/\mathrm{Tr}e^{-\beta \hat{\mathcal{H}}}$, where $\mathrm{Tr}$ denotes the trace in a computational basis.
In the quantum system, the average over the posterior distribution can be replaced with the density matrix as $\mathrm{Tr}(O\hat{\rho})$, where $O$ represents observables such as the training error and the overlap.
The generalization error can be computed from Eq.\eqref{eq7} with the overlap the same as the classical case.

\section{\label{sec:sec3}Mean field analysis}
The typical behaviour of order parameters, such as generalisation error, can be computed from the free energy.
We drive the free energy at the limit of $N,p\rightarrow \infty$ while fixing the number of data per coupling as $\alpha\equiv p/N=O(1)$. 
The free energy density $f$ can be obtained from the partition function $Z=\mathrm{Tr}\exp\left(-\beta \hat{\mathcal{H}}\right)$ as $-\beta f=\lim_{N\rightarrow \infty} 1/N[\ln Z]_{\mathcal{D}}$.
We can not directly apply the replica-calculation to the quantum system due to the non-commutativity  of the spin operator. 
At first, we employ the Suzuki-Trotter decomposition \cite{suzuki_1976} to the following partition function:
\begin{widetext}
\begin{align}
Z&=\lim_{M\rightarrow \infty}\mathrm{Tr}\left\{\exp\left(-\frac{\beta}{M}E(\bm{\hat{\sigma}^z})\right)\exp\left(\frac{\beta \Gamma}{M}\sum_{i=1}^N\hat{\sigma}_i^x \right)\right\}^M\nonumber\\
&=\lim_{M\rightarrow \infty}Z_M,
\label{eq11}
\end{align}
where
\begin{align}
Z_M=&\mathrm{Tr}\prod_{t=1}^M\left(\frac{1}{2}\sinh\left(\frac{2\beta \Gamma}{M}\right)\right)^{\frac{N}{2}}\exp\left\{-\frac{\beta}{M}\sum_{\mu=1}^{p}\Theta\left(-y_{\mu}\mathrm{sgn}\left(\frac{1}{\sqrt{N}}\sum_{i=1}^Nx_{i\mu}{\sigma}_{i}(t)\right)\right)+\frac{1}{2}\ln \coth\left(\frac{\beta\Gamma}{M}\right)\sum_{i=1}^N\sigma_{i}(t)\sigma_{i}(t+1)\right\},
\label{eq12}
\end{align}
the symbol $t$ is the index of the Trotter slice, and $M$ is the Trotter number.
We also impose the periodic boundary conditions $\sigma_i(1)=\sigma_i(M+1)$ for all $i$.
The spin operator is replaced with the classical spin configuration $\sigma_{i}(t)\in{\pm1}$.
In this case, the $\mathrm{tr}$ denotes the summation over all possible spin configurations. 
Since the direct computation of $[\ln Z]_{\mathcal{D}}$ is difficult, we take the configuration average for the replicated system with the replica method \cite{replica_method} as follows:
\begin{align}
[\ln Z]_{\mathcal{D}}=\lim_{n\rightarrow0}\frac{[Z^n]_{\mathcal{D}}-1}{n}.
\label{eq13}
\end{align}
 The replicated partition function is written as 
\begin{align}
\left[Z^n\right]_{\mathcal{D}}&=\lim_{M\rightarrow\infty}\int d\mathcal{D} P(\mathcal{D}|\bm{w})\int d\bm{w}P(\bm{w}) \mathrm{Tr}\left(\frac{1}{2}\sinh\left(\frac{2\beta \Gamma}{M}\right)\right)^{\frac{nMN}{2}}\nonumber\\
&\times\exp\left\{-\frac{\beta}{M}\sum_{a=1}^n\sum_{t=1}^M\sum_{\mu=1}^p\Theta\left(-y_{\mu}\mathrm{sgn}\left(\frac{1}{\sqrt{N}}\sum_{i=1}^Nx_{i\mu}{\sigma}_{ia}(t)\right)\right)+\frac{1}{2}\ln \coth\left(\frac{\beta\Gamma}{M}\right)\sum_{a=1}^n\sum_{t=1}^M\sum_{i=1}^N\sigma_{ia}(t)\sigma_{ia}(t+1)\right\},
\label{eq14}
\end{align}
where $a$ denotes the replica index.
After performing the configuration average, we introduce the order parameters via the delta function as follows:
\begin{align}
m_a(t)&=\frac{1}{N}\sum_{i=1}^Nw_{i}\sigma_{ia}(t), \hspace{5pt}q_{ab}(t,t')=\frac{1}{N}\sum_{i=1}^N\sigma_{ia}(t)\sigma_{ib}(t')\hspace{3pt}(a<b),\hspace{5pt}R_{a}(t,t')=\frac{1}{N}\sum_{i=1}^N\sigma_{ia}(t)\sigma_{ia}(t')\hspace{3pt}(t\neq t')\label{eq15}.
\end{align}
The order parameters can be interpreted as follows: $m_a (t)$ is the magnetisation, $q_{ab}(t,t')$ is the spin-glass order parameter,and $R_{a}(t,t')$ is the correlation between each Trotter slice. 
In addition, we introduce their auxiliary parameters $\tilde{m}_a(t), \tilde{q}_{ab}(t,t'), \tilde{R}_{a}(t,t')$ for the order parameters via the Fourier integral representation of the delta function. 
Under the replica symmetry (RS) ansatz  and static approximation, we have $m_a(t)=m, q_{ab}(t,t')=q,R_{a}(t,t') = R, \tilde{m}_a(t)=\tilde{m}, \tilde{q}_{ab}(t,t')=\tilde{q}, \tilde{R}_{a}(t,t') = \tilde{R}$. 

The free energy density is given by
 \begin{align}
 -\beta f_{\mathrm{RS}}&=\underset{\substack{m,q,R\\ \tilde{m},\tilde{q},\tilde{R}}}{\mathrm{extr}}\left[2\alpha \int DuH\left(-X_1\right) \ln \int D\nu \exp\left(-\beta H\left(X_2\right)\right)+\int Dz \ln 2Y-m\tilde{m}-\frac{1}{2}R\tilde{R}+\frac{1}{2}q\tilde{q}\right]\label{eq16},\\
 X_1&\equiv\sqrt{\frac{m^2}{q-m^2}}u,\hspace{4pt}X_2\equiv \frac{\sqrt{q}u+\sqrt{R-q}\nu}{\sqrt{1-R}},Y\equiv \int Dy \cosh u,\nonumber\\
u&\equiv \sqrt{g^2+(\beta\Gamma)^2},g\equiv\tilde{m}+\sqrt{\tilde{q}}z+\sqrt{\tilde{R}-\tilde{q}}y\label{eq17},
    \end{align}
      where $Du$ denotes the Gaussian measure as $Du=1/\sqrt{2\pi} du e^{-u^2/2}$. Here, we utilise the function $H(x)=\int_x^{\infty}Du$. 
      The order parameters and their auxiliary parameters are determined by the saddle-point condition of the free energy density.
  Detailed calculations for the derivation of Eq.\eqref{eq16} are provided in Appendix \ref{appendix_a}.
The extremisation of Eq.\eqref{eq16} gives rise to the following saddle-point equations:
\begin{align}
m&=\int DzY^{-1}\int Dy\left(\frac{g}{u}\right)\sinh u,\label{eq18}\\
q&=\int Dz\left\{Y^{-1}\int Dy\left(\frac{g}{u}\right)\sinh u\right\}^2,\label{eq19}\\
R&=\int DzY^{-1}\int Dy\left\{\left(\frac{(\beta \Gamma)^2}{u^3}\right) \sinh u+\left(\frac{g}{u}\right)^2\cosh u\right\},\label{eq20}\\
\tilde{m}&=\frac{2\alpha q}{\sqrt{(q-m^2)^3}}\int uDuG\left(X_1\right)\ln \int D\nu \exp\left(-\beta H\left(X_2\right)\right),\label{eq21}\\
\tilde{q}&=\frac{2\alpha m}{\sqrt{(q-m^2)^3}}\int uDuG\left(X_1\right)\ln \int D\nu \exp\left(-\beta H\left(X_2\right)\right)\nonumber\\
&+\frac{2\alpha \beta}{\sqrt{1-R}}\int DuH\left(-X_1\right)\frac{\int D\nu \exp\left(-\beta H\left(X_2\right)\right)G\left(X_2\right)\left\{\frac{\nu}{\sqrt{R-q}}-\frac{u}{\sqrt{q}}\right\}}{\int D\nu \exp\left(-\beta H\left(X_2\right)\right)},\label{eq22}\\
\tilde{R}&=\frac{2\alpha \beta}{\sqrt{(R-q)(1-R)^3}}\int DuH\left(-X_1\right)\frac{\int D\nu \exp\left(-\beta H\left(X_2\right)\right)G\left(X_2\right)\left\{(1-q)\nu+\sqrt{q(R-q)}u\right\}}{\int D\nu \exp\left(-\beta H\left(X_2\right)\right)},\label{eq23}
\end{align}
where we utilise $G(x)=e^{-x^2/2}/\sqrt{2\pi}$. 
The training error is written as 
\begin{align}
\epsilon_t=2\int DuH(-X_1)\frac{\int D\nu H\left(X_2\right)\exp\left(-\beta H\left(X_2\right)\right)}{\int D\nu \exp\left(-\beta H\left(X_2\right)\right)}.
\label{eq24}
\end{align}

In the binary perceptron, the freezing behavior happens in the low temperature \cite{Krauth_1989,Seung_1992}.
This corresponds to the replica symmetry breaking (RSB).
The freezing location can be computed from the zero-entropy condition
as 
\begin{align}
\mathcal{S}&=-\frac{\partial f_{\mathrm{RS}}}{ \partial T}=\beta\left(\alpha \epsilon_t- f_{\mathrm{RS}}\right)=0.
\label{eq25}
\end{align}
\end{widetext}

\begin{figure*}[t]
\centering
\subfigure[\label{fig:fig_1a}]{\includegraphics[width=68mm]{{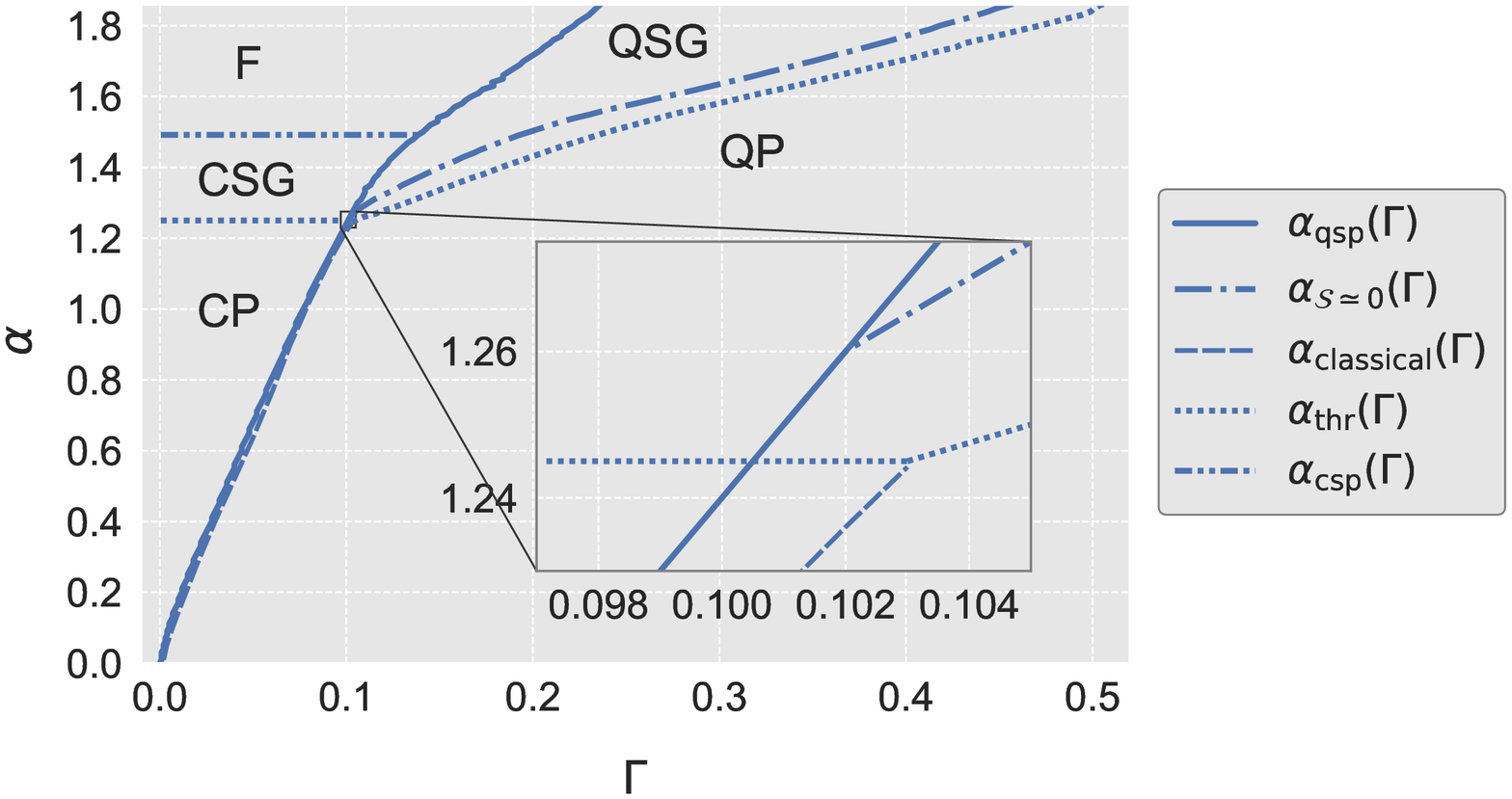}}}
\subfigure[\label{fig:fig_1b}]{\includegraphics[width=55mm]{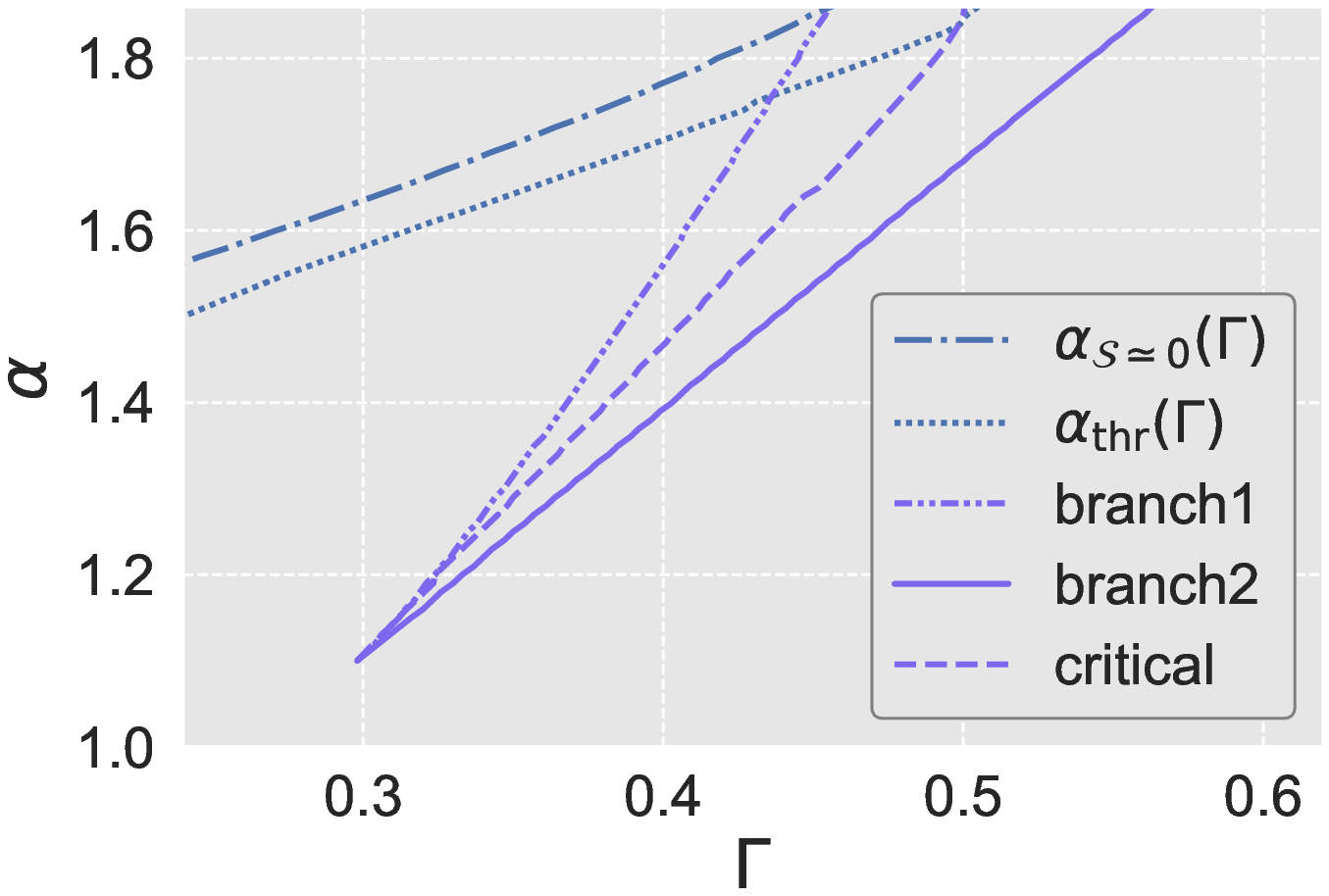}}
\caption{(a) Phase diagram of the binary perceptron with quantum fluctuations. 
The horizontal axis represents the strength of the transverse field.
The vertical axis denotes the pattern ratio. 
The inset shows the neighborhood where  the $\alpha_{\mathrm{thr}}(\Gamma)$ and $\alpha_{\mathrm{classical}}(\Gamma)$ curves intersect.
(b) The spinodal curves and the ``critical'' one in $\alpha>1.0$. 
The $\alpha_{\mathcal{S}\simeq0}(\Gamma)$ and  $\alpha_{\mathrm{thr}}(\Gamma)$ curves are the same as those in Fig.\ref{fig:fig_1a}.
}
\label{fig:fig_1}
\end{figure*}

\section{\label{sec:sec4}Numerical experiments}
We numerically solve the saddle-point equations in Eqs.\eqref{eq18}-\eqref{eq23} using the inverse temperature $\beta=20$.
A phase diagram of the binary perceptron with quantum fluctuations is presented in Fig.\ref{fig:fig_1a}.
Since the behaviors of the saddle-point equations becomes unstable in $\alpha>1.87$, we consider $\alpha \leq1.86$. 
At first, we briefly outline the classical model ($\Gamma=0$) in Eq.\eqref{eq4} and move on to the quantum model ($\Gamma\neq0$) in Eq.\eqref{eq9}.
For the classical model, 
the review of the previous study \cite{Seung_1992} is included.

In the binary perceptron, two solutions, one with the perfect generalization ($m=1$) and another with the poor generalization ($m<1$), exist.
In the lower pattern ratio than $\alpha_c\simeq 1.245$, the $m<1$ solution has the lowest free energy.
The $m=1$ solution is the metastable state. 
We denote the phase as the ``Classical Paramagnetic'' (CP) phase.
As we increase the pattern ratio, the generalization performance of the model is increased. 
When the pattern ratio reaches $\alpha =\alpha_c$, the first-order phase transition to the $m=1$ solution happens.
We denote the thermodynamic transition point as $\alpha_{\mathrm{thr}}(\Gamma=0)=\alpha_c$.
For $\alpha>\alpha_{c}$, the $m=1$ solution has the lowest free energy.
The $m<1$ solution is the metastable state and has a negative entropy beyond the zero-entropy point $\alpha_{\mathcal{S}\simeq0}(\Gamma=0)$.
This implies that the replica symmetry breaking (RSB) occurs and  the RS ansatz is invalid for the metastable $m<1$ solution.
Note that the thermodynamic transition and the RSB coincide almost simultaneously at $T\simeq 0$ and $\alpha_{\mathcal{S}\simeq0}(\Gamma=0)\simeq \alpha_c$.
At $T=0$, two transitions happen at the same time.
Since $\alpha_{\mathcal{S}\simeq0}(\Gamma=0)$ can not be distinguished with $\alpha_c$ in Fig.\ref{fig:fig_1a}, we omit $\alpha_{\mathcal{S}\simeq0}(\Gamma=0)$ in the classical case.
Until the spinodal point $\alpha_{\mathrm{csp}}(\Gamma=0)\simeq 1.492$, the metastable $m<1$ solution remains.
We denote the phase between $\alpha_{\mathcal{S}\simeq0}(\Gamma=0)\simeq \alpha_c$ and $\alpha_{\mathrm{csp}}(\Gamma=0)$  as the ``Classical Spin Glass'' (CSG) phase \footnote{Strictly speaking, the CSG phase is characterized by $m=0$ and $q>0$ in the SG theory.
Following Ref.\cite{Seung_1992}, we adopt the word as the ``SG'' phase  to represent the RSB for the metastable state.
}.
Beyond $\alpha_{\mathrm{csp}}(\Gamma=0)$, the metastable $m<1$ solution disappears and the $m=1$ solution only exists.
We denote the phase as the ``Ferromagnetic'' (F) phase.
To describe the correct boundary between the SG phase and the F phase, we need to construct the RSB solutions.
The vanishing point of the metastable $m<1$ solution changes from $\alpha_{\mathrm{csp}}(\Gamma=0)=1.492$ to $\alpha_{\mathrm{csp}}(\Gamma=0)=1.628$ under the 1-step RSB (1RSB) ansatz \cite{Seung_1992}.
Since the equilibrium state is the $m=1$ solution for $\alpha > \alpha_c$, the thermodynamics transition point is valid under the RS ansatz.

We extend the above discussions into the quantum model.
The quantum model has the solution with the poor generalization ($m<1$) in the high strength of the transverse field.
The $m=1$ solution and the classical $m<1$ solution remains under the transverse field.
Below $\alpha_c$, the first-order phase transition between the quantum $m<1$ solution and the classical $m<1$ solution happens at the $\alpha_{\mathrm{classical}}(\Gamma)$ curve.
On the left side of the $\alpha_{\mathrm{classical}}(\Gamma)$ curve, the classical $m<1$ solution has the lowest free energy.
The $m=1$ solution and the quantum $m<1$ solution are the metastable states. 
The phase is represented by the CP phase.
On the right side of the $\alpha_{\mathrm{classical}}(\Gamma)$ curve,
the quantum $m<1$ solution has the lowest free energy.
We denote the phase as the ``Quantum Paramagnetic'' (QP) phase.
In the QP phase, the $m=1$ solution and the classical $m<1$ solution are the metastable states.
At the $\alpha_{\mathrm{qsp}}(\Gamma)$ curve, the freezing phenomenon to the classical model occurs and the quantum $m<1$ solution vanishes.
Above the $\alpha_{\mathrm{qsp}}(\Gamma)$ curve, quantum fluctuations do not affect the system, and the model is identical to the classical model.
The structure of the phase under the transverse field inherits the classical model.
At  $\alpha=0$, we can attain $m=q=0$ and $R=\tanh(\beta \Gamma)/(\beta \Gamma)$   from $\tilde{m}=\tilde{q}=\tilde{R}=0$.
Since $R\rightarrow 0$ holds in the limit of $\Gamma\rightarrow 0$, the  $\alpha_{\mathrm{qsp}}(\Gamma)$ curve passes thorough origin.
In the QP and CP phases, the free energy densities are given by  $f_{\mathrm{QP}}(\alpha=0)=-T\ln2\cosh(\beta \Gamma)$ and $f_{\mathrm{CP}}(\alpha=0)=-T\ln2$.
The inequality $f_{\mathrm{QP}}\leq f_{\mathrm{CP}}$ holds under the transverse field.
Since the equality holds in the limit of $\Gamma\rightarrow0$, the $\alpha_{\mathrm{classical}}(\Gamma)$ and $\alpha_{\mathrm{qsp}}(\Gamma)$ curves
intersect the origin.
Beyond $\alpha_c$, the thermodynamic transition to the $m=1$ solution happens at the $\alpha_{\mathrm{thr}}(\Gamma)$ curve.
In the quantum model, the thermodynamic transition point depends on the strength of the transverse field. 
The $\alpha_{\mathrm{thr}}(\Gamma)$ and  $\alpha_{\mathrm{classical}}(\Gamma)$ curves intersect around $\Gamma\simeq 0.103$ and $\alpha\simeq\alpha_c$.
Above the $\alpha_{\mathrm{thr}}(\Gamma)$ curve, the $m=1$ solution has the lowest free energy  under the transverse field.
Above the $\alpha_{\mathcal{S}\simeq0}(\Gamma)$ curve, the quantum $m<1$ solution has a negative entropy and the RSB takes place.
We denote the phase as the ``Quantum Spin Glass'' (QSG) phase.
Like the classical model, the RS ansatz is invalid to describe the metastable state and is valid for the equilibrium state.
To attain the correct boundary between the F phase and the QSG phase, we need to consider the RSB solutions.
Between the $\alpha_{\mathcal{S}\simeq0}(\Gamma)$ and  $\alpha_{\mathrm{thr}}(\Gamma)$ curves, the RSB does not occur and the equilibrium state is the $m=1$ solution \footnote{This phase corresponds to the ``Metastability''  as denoted in Ref.\cite{Seung_1992}}.

Figure.\ref{fig:fig_1b} shows the spinodal curves of the two different quantum $m<1$ solutions.
We denote these curves as the ``branch 1'' and ``branch 2'' curves.
At the ``critical'' curve, the first-order phase transition occurs and the two solutions have the same free energy.
The $\alpha_{\mathrm{thr}}(\Gamma)$ and $\alpha_{\mathcal{S}\simeq0}(\Gamma)$ curves are the same as those in Fig.\ref{fig:fig_1a}.
Under the ``critical'' curve, the ``branch 1'' solution has the lowest free energy.
On the other hand, the ``branch 2'' solution has the lowest free energy above the ``critical'' curve. 
Around $\alpha=1.85$ and $\Gamma=0.5$, the``critical'' curve intersects the $\alpha_{\mathrm{thr}}(\Gamma)$ curve.
Above the intersection,
the $m=1$ solution has the lowest free energy and the ``branch 2'' solution is always the metastable state.

\begin{figure}[t]
\includegraphics[width=72mm]{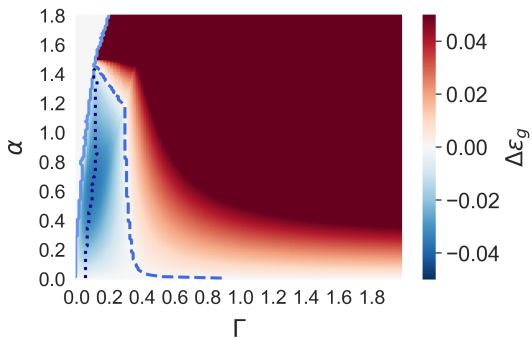}
\caption{Heat map of the differences in generalisation errors $\Delta\epsilon_g$ between the quantum model and classical model.
Both axes are the same as those in Fig.\ref{fig:fig_1}.
The solid curve denotes the $\alpha_{\mathrm{qsp}}(\Gamma)$ curve. 
The dashed curve represents the location of $\Delta\epsilon_g\simeq0$.
The dotted curve is the location of minimum $\Delta\epsilon_g$.
}
\label{fig:fig_2}
\end{figure}

Figure.\ref{fig:fig_2} presents a heat map of the differences in generalisation error between the quantum model and classical model as $\Delta\epsilon_g=\epsilon_g-\epsilon_g^{\mathrm{classical}}$.
The dashed curve indicates the location of $\Delta \epsilon_g\simeq 0$.
The solid curve denotes the $\alpha_{\mathrm{qsp}}(\Gamma)$ curve in Fig.\ref{fig:fig_1a}.
At the dashed curve, the strength of quantum fluctuations corresponds to the strength of thermal fluctuation of the classical model.
Above the dashed curve, the generalisation error of the quantum model is greater than that of the classical model.
Under the dashed curve, quantum fluctuations enhance the generalisation performance of the binary perceptron.
The maximum generalisation performance can be attained at the dotted curve.
Learning with the low strength of the transverse field improves the generalization performance.

\begin{figure}[t]
\includegraphics[width=70mm]{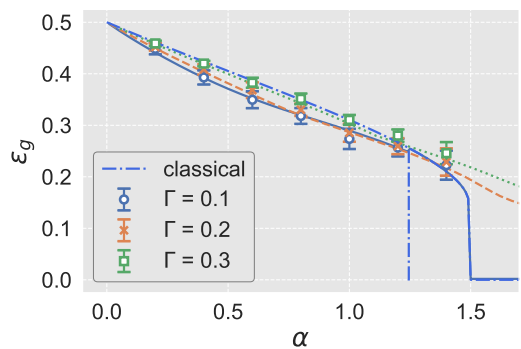}
\caption{Dependence of the generalisation error on the pattern ratio of the fixed strength of the transverse field for $\Gamma=0.1$, $0.2$, and $0.3$.
The lines are derived from the saddle-point equations. 
The symbols represent the results obtained by the quantum Monte Carlo simulations.
The ``classical'' line represents the results of the classical model.}
\label{fig:fig_3}
\end{figure}

To verify the RS solutions, we perform quantum Monte Carlo simulations. 
We set the number of parameters to $N=257$, Trotter number to $M=128$, and temperature to $T=0.05$. 
We calculate a $200000$ Monte Carlo step (MCS) average after $100000$ MCS equilibration for each instance.
We compute the configuration average over the data distribution by randomly generating $20$ instances.
The results obtained from the quantum Monte Carlo simulations represent the expectations of all Trotter slices.
In Fig.\ref{fig:fig_3}, we present the behaviours of the generalisation error in terms of the pattern ratio of the fixed strength of the transverse field for $\Gamma=0.1$, $0.2$, and $0.3$.
The error bar represents the standard deviation. 
The numerical results are consistent with the RS solutions, except for the high pattern ratio.
One can see that quantum fluctuations improve the generalisation performance of the binary perceptron compared to the classical model.
As the pattern ratio increases, the energy landscape becomes increasingly non-convex and the local spin-flip algorithm easily becomes stuck in local minima.
Therefore, it is difficult to estimate the order parameters in this region.

\begin{figure}[t]
\includegraphics[width=60mm]{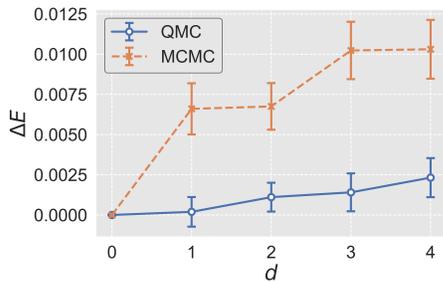}
\caption{Training error differences between the reference solutions obtained by the quantum Monte Carlo simulations and Markov-Chain Monte Carlo simulations, and the neighbouring solutions, whose hamming distances from the reference solutions are denoted by $d$.
The symbols are averaged over all neighbouring solutions within the hamming distance $d$. }
\label{fig:fig_4}
\end{figure}

We investigate the robustness of the solutions obtained by the quantum Monte Carlo simulations based on the energy landscape around the solutions.
We utilised the final results of the quantum Monte Carlo simulations as $\hat{\sigma}_i=\mathrm{sgn}(\sum_{t=1}^M\sigma_{it}) \quad (i=1,...,N)$ for $\alpha=0.4$ and $\Gamma=0.1$ in Fig.\ref{fig:fig_3}. 
We call these solutions ``reference solutions''. 
We calculate the local training error differences between the reference solutions and neighbouring solutions, whose Hamming distances from the reference solutions are denoted by $d$.
In Fig.\ref{fig:fig_4}, we plot the local training error differences between the reference solutions and neighbouring solutions.
We compute the average differences over all neighbouring solutions.
The error bar represents the standard deviation.
For comparison, we also plot the results of the reference solutions of the classical model attained by Markov-Chain Monte Carlo simulations. 
The experimental settings and instances are the same as those shown in Fig.\ref{fig:fig_3}.
The local training error differences are related to the flatness of the energy landscape. 
The training error of the flat minimiser is robust to perturbations.
One can see that the local training error differences of the quantum model are less than those of the classical model.
Overall, quantum fluctuations lead solutions toward the flat minimiser. 

\begin{figure}[t]
\subfigure[\label{fig:fig_5a}]{\includegraphics[width=42mm]{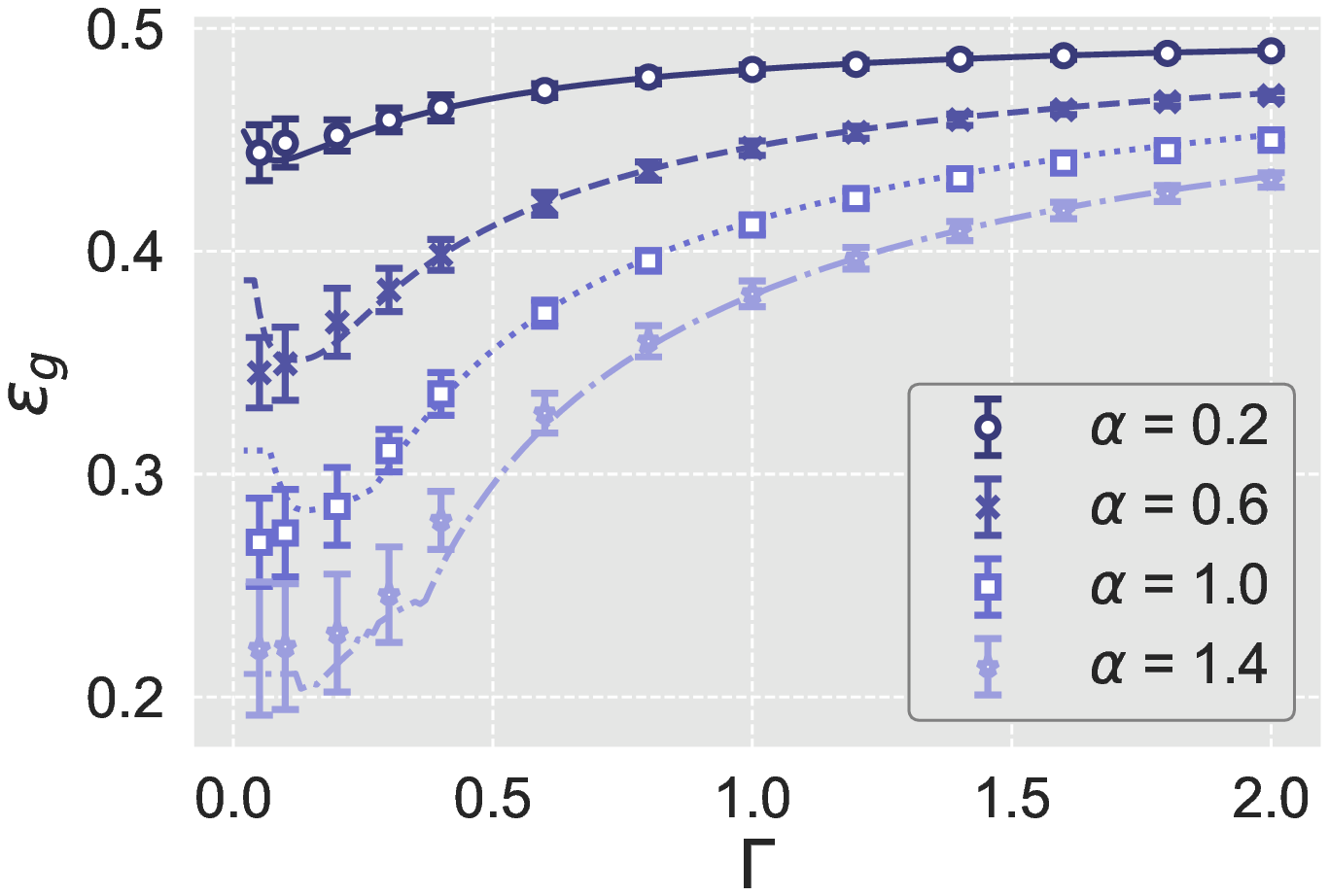}}
\subfigure[\label{fig:fig_5b}]{\includegraphics[width=42mm]{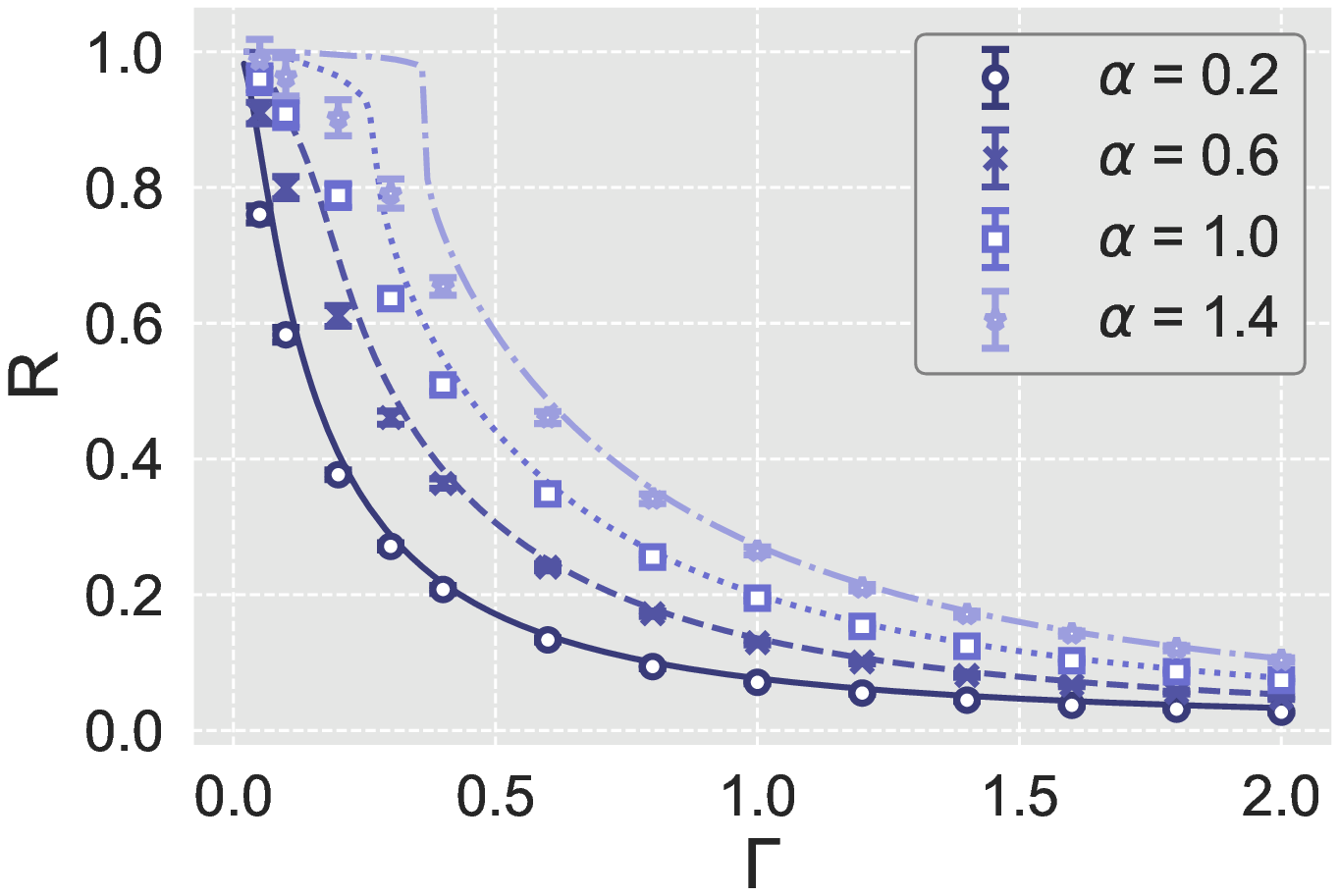}}
\caption{Dependence of order parameters on the strength of the transverse field for fixed pattern ratios of $\alpha=0.2,0.6,1.0$ and $1.4$.
The vertical axes denote the following order parameters: (a) generalisation error and (b) correlation between Trotter slices. 
Each line is derived from the saddle-point equations. 
The symbols are derived from the quantum Monte Carlo simulations.
 }
\label{fig:fig_5}
\end{figure}

Finally, we plot the behaviours of the generalisation error and the correlations between Trotter slices with respect to the strength of the transverse field for the fixed pattern ratios of $\alpha=0.2, 0.6$, $1.0$, and $1.4$ in Fig.\ref{fig:fig_5}. 
The numerical results are consistent with the RS solutions, except for the low strength of the transverse field and high pattern ratio.
The deviation between the RS solutions and numerical results stems from two factors.
The first factor is the violation of ergodicity. 
The energy landscape of the binary perceptron contains many local minima.
By introducing quantum fluctuations, the energy landscape of the quantum model becomes smoother than that of the classical model.
Therefore, the sampling of spin configurations from the Gibbs-Boltzmann distribution of the quantum model is easier than that in the classical model.
With a high strength of the transverse field and low pattern ratio, the local spin-flip algorithm works well and does not become stuck in local minima.
As the strength of the transverse field decreases, the smoothness of the energy landscape decreases.
In this case, the local spin-flip algorithm becomes stuck in local minima due to the non-convexity of the energy landscape.
For a high pattern ratio, the deviation between the RS solutions and numerical results is greater than that for a low pattern ratio.
The second factor is the violation of the static approximation. 
To derive saddle-point equations, we adopt the static approximation.
Generally, the order parameters depend on the Trotter slices.
For the generalisation error, deviations between the RS solutions and numerical results are smaller than those for the correlation between Trotter slices.
In this case, the effects of violating the static approximation are small.
The RS solutions for the generalisation error are valid, except for the low strength of the transverse field.
Because the correlation between Trotter slices depends on each Trotter slice, the RS solutions are invalid, except for the high strength of the transverse field.
A similar behaviour occurs in the code-division multiple-access model \cite{arai2020mean}.
In general, to verify the validity of the static approximation is difficult except for the limited model \cite{Obuchi_2007}.

\section{\label{sec:sec5}Conclusion}

We analysed the teacher-student learning of the binary perceptron with quantum fluctuations.
The energy landscape of the binary perceptron contains many local minima.
Standard SA often becomes stuck in local minima and fails to identify the ground state of the system. 
By introducing quantum fluctuations, the state evolves more efficiently because the effective energy landscape becomes smoother. 
Additionally, quantum fluctuations improve the robustness of solutions characterised by wide-flat minima.
We determined that these robust solutions yield better generalisation performance than the dominated solutions identified by standard SA.

We showed the phase diagram of the binary perceptron with quantum fluctuations. 
In a weak transverse field and low pattern ratio,
the first-order phase transition between the $m<1$ solutions of the classical and quantum models occurs at the $\alpha_{\mathrm{classical}}(\Gamma)$ curve.
On the left side of the $\alpha_{\mathrm{qsp}}(\Gamma)$ curve, the freezing behavior to the classical model happens, and quantum fluctuations do not affect the system.
In this case, the model is identical to the classical model.
The $\alpha_{\mathrm{thr}}(\Gamma)$ curve denotes the thermodynamics transition point between the $m=1$ solution and the $m<1$ solutions.
In a strong transverse field, we need more data to do perfect generalisation.
This phenomenon occurs in the classical model at a high temperature.
We also showed the first-order phase transition between quantum $m<1$ solutions from different branches under a transverse field. 
The quantum $m<1$ solution has the RSB at the $\alpha_{\mathcal{S}\simeq0}(\Gamma)$ curve.
The RSB for the metastable state happens in the classical model. 
In the classical model, this RSB is characterised by the frozen-1RSB \cite{Krauth_1989,Seung_1992,Aubin_2019a}.
The frozen-1RSB was discovered in the random energy model \cite{Derrida1981}.
The zero-entropy condition can be derived from the 1RSB ansatz and take the limit of the inner cluster overlap and its conjugate parameter as $q_1\rightarrow 1$  and $\tilde{q}_1\rightarrow \infty$.
The different cluster overlap $q_0$ is finite.
We expect that the quantum model has the same structure as the classical model. 
To investigate the frozen-1RSB structure in detail,
we need the 1RSB solutions for the quantum model.
The quantum random energy model \cite{Goldschmidt1990,Obuchi_2007} has the similar frozen-1RSB structure in the SG phase.
The 1RSB solutions for the quantum model can be considered.
For the correlation between each Trotter slice, the inner cluster overlap, and the different cluster overlap, 
the relationship holds as $R\geq q_1\geq q_0$.
In the frozen-1RSB, $R=q_1\rightarrow1$ holds.
This implies that the $\alpha_{\mathcal{S}\simeq0}$ curve matches  the $\alpha_{\mathrm{qsp}}$ curve.
For $\alpha_c< \alpha< 1.628$ above the $\alpha_{\mathcal{S}\simeq0}$ curve, the phase is characterised by the CSG phase.
For $\alpha \geq  1.628$ above the $\alpha_{\mathcal{S}\simeq0}$ curve, the phase is represented by the F phase.
In Fig.\ref{fig:fig_1}, we denote the QSG phase characterised by $R\neq1$ and $\mathcal{S}<0$.
In the frozen-1RSB structure, this phase does not exist.
The validation of above conjecture for the frozen-1RSB  will be done in a future study.

We showed a heat map of the differences in generalisation error between the quantum model and classical model. 
In a strong transverse field, the generalisation performance of the binary perceptron is worse than that of the classical model. 
In a weak transverse field, quantum fluctuations enhance the generalisation performance of the binary perceptron. 
These results were validated by quantum Monte Carlo simulations.
By introducing quantum fluctuations, spin configurations can be efficiently sampled from the posterior distribution of the weight parameter.
This implies that the energy landscape of the quantum model is much smoother than that of the classical model.
To investigate the energy landscape in detail,
we checked the robustness of the reference solutions quantum Monte Carlo simulations.
We calculated the average local training error differences between the reference solutions and all neighbouring solutions.
The local training error differences between the reference solutions and neighbouring solutions obtained by quantum Monte Carlo simulations were smaller than those obtained by the Markov-chain Monte Carlo method.
Finally we conclude that quantum fluctuations bring states to the flat minimiser. 

The term introduced by the transverse field can explain why quantum fluctuations lead to he solutions in the flat valley.
The transverse field term yields the interaction between the Trotter slices.
In a strong transverse field, the system in each Trotter slice with the effective inverse temperature $\beta/M$ minimizes the classical energy term.
In this case, the solutions are the dominant states in the classical model, which are in the sharp valley.
On the other hand, a large ferromagnetic interaction between the Trotter slices emerges in a small transverse field.
The transverse field term induces the spin configurations in each Trotter slice closer together and assigns the lower energies to the solutions in the flat valley.
As discussed in Ref.\cite{Chaudhari_2019}, the solutions in the flat valley have better generalisation performance.
The flat minimiser can be reached algorithmically by exploiting the local entropy.
In the robust ensemble (RE) model \cite{Baldassi_2016b}, the cost function can be derived from maximising the local entropy.
As explained in Ref.\cite{Baldassi_2017}, the cost functions of the RE model and the effective classical model in Eq.\eqref{eq12} are similar.
Interestingly, quantum fluctuations induce the solutions toward the flat minimiser not artificially but naturally.
In addition, the ferromagnetic interaction term between each Trotter slice can be interpreted as the $L_2$ norm regularisation term between each Trotter slice as $||\bm{\sigma}(t)-\bm{\sigma}(t')||^2_2\propto-2\sum_{i}\sigma_i(t)\sigma_i(t')$.
In the context of machine learning, the regularisation term prevents overfitting and enhances the generalisation performance of the model.
Quantum fluctuations suppress overfitting and improve the generalisation performance of the model naturally.

In this study, we considered a single-layer perceptron. 
In the future, our results can be extended to a multilayer perceptron \cite{Schwarze_1992,Schwarze_1993a,Schwarze_1993b,Monasson_1995,Aubin_2019b}.
The multilayer perceptron contains a high local entropy region \cite{Baldassi_2016b}. 
Therefore, we expect that quantum fluctuations will also improve the generalisation performance of the multilayer perceptron.
We can also consider a rotationally invariant model \cite{Kabashima_2008,Shinzato_2008a,Shinzato_2008b}. 
In the classical model, orthogonal input data increase the critical capacity.
Determining whether quantum fluctuations enhance the generalisation performance for different types of datasets is another interesting topic for future study.

\section*{Acknowledgments}
M.O. was supported by KAKENHI (No. 19H01095), and the Next Generation High-
Performance Computing Infrastructures and Applications R $\&$ D Program by MEXT.
K.T. was supported by JSPS KAKENHI (No. 18H03303). This work was partially supported by JST-CREST (No. JPMJCR1402). 

\appendix
\begin{widetext}
\section{DERIVATION OF FREE ENERGY
\label{appendix_a}}
We derive the free energy density under the RS ansatz and static approximation.
We introduce following terms: 
\begin{align}
u_{a}^\mu(t)&=\frac{1}{\sqrt{N}}\sum_{i=1}^N{x}_{i\mu}\sigma_{ia}(t),\label{A1}\\
u_{0}^{\mu}&=\frac{1}{\sqrt{N}}\sum_{i=1}^Nx_{i\mu}w_{i}.\label{A2}
\end{align}
The energy function term in Eq.\eqref{eq10} can be expressed as 
\begin{align} \exp\left\{-\frac{\beta}{M}\sum_{a=1}^n\sum_{t=1}^M\sum_{\mu=1}^{p}\Theta\left(-y_{\mu}\mathrm{sgn}\left(\frac{1}{\sqrt{N}}\sum_{i=1}^Nx_{i\mu}{\sigma}_{ia}(t)\right)\right)\right\} &=\prod_{a=1}^n\prod_{t=1}^M\prod_{\mu=1}^{\alpha N}\exp\left(-\frac{\beta}{M}\Theta\left(-u_0^{\mu}u_{a}^{\mu}(t)\right)\right)\nonumber \\ 
     &=\prod_{a=1}^n\prod_{t=1}^M\prod_{\mu=1}^{\alpha N}\left\{\exp\left(-\frac{\beta}{M}\right)+\Theta\left(u_0^{\mu}u_{a}^{\mu}(t)\right)\left(1-\exp\left(-\frac{\beta}{M}\right)\right)\right\}.\label{A3}
\end{align}

When $x_{i\mu}$ are i.i.d. random variables with zero mean and unit variance, the central limit theorem guarantees that $u_{a}^\mu(t)$ and $u_{0}^\mu$ follow the multivariate Gaussian random variables, which are characterised by zero mean and covariances of $[u_{a}^{\mu}(t)u_{b}^{\upsilon}(t')]_{\mathcal{D}}=Q_{ab}(t,t')\delta_{\mu \upsilon}$ for a fixed parameter of $\bm{w},\{\bm{\sigma}_{t}^z(a)\} (a=1,\dots,n,t=1,\dots,M)$ \cite{Kabashima_2009}.
To simplify notation, we utilise the representation $u_{0}^{\mu}(t)=u_0^\mu$. The covariance matrix can be written as  
\begin{align}
Q_{ab}(t,t')&\equiv\begin{cases}
q_{ab}(t,t') &\quad (a>b=1,\dots,n;b>a=1,\dots,n;t,t'=1,\dots,M )\\
R_{a}(t,t') &\quad (a=b=1\dots,n;t\neq t')\\
m_a(t) &\quad (a=0,b=1,\dots,n;a=1,\dots,n, b=0;t=1,\dots, M).\\
1 &\quad (a=b=0\hspace{5pt}\mathrm{or}\hspace{5pt} a=1,\dots,n;t=1,\dots,M) 
\end{cases}.\label{A4}
\end{align}
The integration of the data distribution $P(\mathcal{D}|\bm{w})$ can be replaced by the multivariate Gaussian distribution $P(u_0,\{u_{a}(t)\})$.
Next, we introduce the delta function and its Fourier integral representation as follows:
\begin{align}
&\prod_{a,t}\int dm_a(t)\delta\left(m_a(t)-\frac{1}{N}\sum_{i=1}^Nw_i\sigma_{ia}(t)\right)\nonumber\\
&=\prod_{a,t}\int \frac{iNdm_a(t)d\tilde{m}_a(t)}{2\pi M}\exp\left\{-\frac{\tilde{m}_a(t)}{M}\left(Nm_a(t)-\sum_{i=1}^Nw_i\sigma_{ia}(t)\right)\right\},\label{A5}\\
&\prod_{a,t\neq t'}\int dR_{a}(t,t')\delta\left(R_{a}(t,t')-\frac{1}{N}\sum_{i=1}^N\sigma_{ia}(t)\sigma_{ia}(t')\right)\nonumber\\
&=\prod_{a,t\neq t'}\int \frac{iNdR_{a}(t,t')d\tilde{R}_{a}(t,t')}{4\pi M^2}\exp\left\{-\frac{\tilde{R}_{a}(t,t')}{2M^2}\left(NR_{a}(t,t')-\sum_{i=1}^N\sigma_{ia}(t)\sigma_{ia}(t')\right)\right\},\label{A6}\\
&\prod_{a<b,t,t'}\int dq_{ab}(t,t')\delta\left(q_{ab}(t,t')-\frac{1}{N}\sum_{i=1}^N\sigma_{ia}(t)\sigma_{ib}(t')\right)\nonumber\\
&=\prod_{a<b,t,t'}\int \frac{iNdq_{ab}(t,t')d\tilde{q}_{ab}(t,t')}{2\pi M^2}\exp\left\{-\frac{\tilde{q}_{ab}(t,t')}{M^2}\left(Nq_{ab}(t,t')-\sum_{i=1}^N\sigma_{ia}(t)\sigma_{ib}(t')\right)\right\},\label{A7}
\end{align}
Finally, we can rewrite the replicated partition function as 
\begin{align}
\left[Z^n\right]_{\mathcal{D}}&=\lim_{M\rightarrow \infty}\left(\prod_{a,t}\int \frac{iNdm_a(t)d\tilde{m}_a(t)}{2\pi M}\right)\left(\prod_{a,t\neq t'}\int \frac{iNdR_{a}(t,t')d\tilde{R}_{a}(t,t')}{2\pi M^2}\right)\left(\prod_{a<b,t,t'}\int \frac{iNdq_{ab}(t,t')d\tilde{q}_{ab}(t,t')}{2\pi M^2}\right)e^{G_1+G_2+G_3}\label{A8}\\
e^{G_1}&\equiv\exp\left\{\alpha N\ln \left[ \prod_{a,t}\left\{\exp\left(-\frac{\beta}{M}\right)+\Theta\left(u_0^{\mu}u_{a}^{\mu}(t)\right)\left(1-\exp\left(-\frac{\beta}{M}\right)\right)\right\} \right]_{\bm{u}}\right\},\label{A9}\\
e^{G_2}&\equiv \int d\bm{w}P(\bm{w})\mathrm{Tr}\left(\frac{1}{2}\sinh\left(\frac{2\beta \Gamma}{M}\right)\right)^{\frac{nMN}{2}}\exp\left\{\frac{1}{M}\sum_{a,t}\tilde{m}_a(t)\sum_{i=1}^Nw_i\sigma_{ia}(t)+\frac{1}{2M^2}\sum_{a,t\neq t'}\tilde{R}_{a}(t,t')\sum_{i=1}^N\sigma_{ia}(t)\sigma_{ia}(t')\right.\nonumber\\
&\left.+\frac{1}{M^2}\sum_{a<b}\sum_{t,t'}\tilde{q}_{ab}(t,t')\sum_{i=1}^N\sigma_{ia}(t)\sigma_{ib}(t')+\frac{1}{2}\ln \coth\left(\frac{\beta\Gamma}{M}\right)\sum_{a=1}^n\sum_{t=1}^M\sum_{i=1}^N\sigma_{ia}(t)\sigma_{ia}(t+1)\right\},\label{A10}\\
e^{G_3}&\equiv \exp\left\{ -\frac{N}{M}\sum_{a,t}\tilde{m}_a(t)m_a(t)-\frac{N}{2M^2}\sum_{a,t\neq t'}\tilde{R}_{a}(t,t')R_{a}(t,t')-\frac{N}{M^2}\sum_{a<b,t,t'}\tilde{q}_{ab}(t,t')q_{ab}(t,t') \right\}, \label{A11} 
\end{align}
where $[ \cdot]_{\bm{u}}$ represents the expectation over the multivariate Gaussian distribution $P(u_0,\{u_{a}(t)\})$.
We adopt the RS ansatz  and  static approximation as follows:
\begin{align}
m_a(t)&=m,q_{ab}(t,t')=q,R_{a}(t,t')=R,\nonumber \\
\tilde{m}_a(t)&=\tilde{m},\tilde{q}_{ab}(t,t')=\tilde{q}, \tilde{R}_{a}(t,t')=\tilde{R}. 
\label{A12}
\end{align}
Under the RS ansatz and static approximation, the Gaussian random variables can be expressed as 
 \begin{align}
 u_{a}^\mu(t)&=\sqrt{q}u+\sqrt{R-q}\nu_a+\sqrt{1-R}\upsilon_t\label{A13}\quad(a=1,\dots,n;t=1,\dots,M),\\
  u_0^\mu&=\sqrt{\frac{m^2}{q}}u+\sqrt{1-\frac{m^2}{q}}\nu_0,\label{A14}
 \end{align}
 where $u$, $\{\nu_a\}$, and $\{\upsilon_t\}$ are i.i.d. Gaussian random variables with zero mean and unit variance. In Eq.\eqref{A9}, the integration over $P(u_0,\{u_{a}(t)\})$ can be performed as
\begin{align}
&\left[ \prod_{a,t}\left\{\exp\left(-\frac{\beta}{M}\right)+\Theta\left(u_0^{\mu}u_{a}^{\mu}(t)\right)\left(1-\exp\left(-\frac{\beta}{M}\right)\right)\right\} \right]_{\bm{u}}\nonumber\\
&=\int Du\int D\nu_02\Theta(u_0^\mu)\prod_{a=1}^n\int D\nu_a\prod_{t=1}^M\int D\upsilon_t\left\{\exp\left(-\frac{\beta}{M}\right)+\Theta\left(u_{a}^{\mu}(t)\right)\left(1-\exp\left(-\frac{\beta}{M}\right)\right)\right\}\nonumber\\
&=2\int Du H\left(-\sqrt{\frac{m^2}{q-m^2}u}\right)\prod_{a=1}^n\int D\nu_a\prod_{t=1}^M\left\{\exp\left(-\frac{\beta}{M}\right)+H\left(-\frac{\sqrt{q}u+\sqrt{R-q}\nu_a}{\sqrt{1-R}}\right)\left(1-\exp\left(-\frac{\beta}{M}\right)\right)\right\}\nonumber\\
&\simeq \exp\left[2n\int DuH\left(-X_1\right)\ln \int D\nu_a\prod_{t=1}^M\left\{\exp\left(-\frac{\beta}{M}\right)+H\left(-X_2\right)\left(1-\exp\left(-\frac{\beta}{M}\right)\right)\right\}\right]\nonumber\\
&\simeq \exp\left\{2n\int DuH\left(-X_1\right)\ln \int D\nu_a\left[\left\{1-\frac{\beta}{M}\left(1-H\left(-X_2\right)\right)\right\}^{-\frac{M}{\beta\left(1-H\left(-X_2\right)\right)}}\right]^{-\beta\left(1-H\left(-X_2\right)\right)}\right\}\nonumber\\
&=\exp\left[2n\int DuH\left(-X_1\right)\ln \int D\nu_a\exp\left\{-\beta\left(1-H\left(-X_2\right)\right)\right\}\right]\nonumber\\
&=\exp\left[2n\int DuH\left(-X_1\right)\ln \int D\nu\exp\left\{-\beta H\left(X_2\right)\right\}\right],
\label{A15}
\end{align}
where we utilise the relationship $1-H(x)=H(-x)$ and rewrite $\nu_a$ as $\nu$ in the final equation.
We can attain  $e^{G_1}$ by substituting Eq.\eqref{A15} for Eq.\eqref{A9}.

We calculate $e^{G_2}$ under the RS ansatz  and static approximation as follows:
\begin{align}
e^{G_2}&= \int d\bm{w}P(\bm{w})\mathrm{Tr}\int Dz\left(\frac{1}{2}\sinh\left(\frac{2\beta \Gamma}{M}\right)\right)^{\frac{nMN}{2}} \exp\left\{\frac{\tilde{m}}{M}\sum_{a,t,i}w_i\sigma_{ia}(t)+\frac{\tilde{R}-\tilde{q}}{2M^2}\sum_{a,i}\left(\sum_{t=1}^M\sigma_{ia}(t)\right)^2+\frac{\sqrt{\tilde{q}}}{M}\sum_{a,t,i}z\sigma_{ia}(t)\right.\nonumber\\
&\left.+\frac{1}{2}\ln \coth\left(\frac{\beta\Gamma}{M}\right)\sum_{a=1}^n\sum_{t=1}^M\sum_{i=1}^N\sigma_{ia}(t)\sigma_{ia}(t+1)\right\}\nonumber\\
&=\prod_{i=1}^N\sum_{w_{i}=\pm1}\frac{1}{2}\int Dz \prod_{a=1}^n\int Dy \prod_{t=1}^M\mathrm{Tr}\left(\frac{1}{2}\sinh\left(\frac{2\beta \Gamma}{M}\right)\right)^{\frac{1}{2}}\nonumber\\
&\times \exp\left\{\frac{1}{M}\left(w_i\tilde{m}+\sqrt{\tilde{q}}z+\sqrt{\tilde{R}-\tilde{q}}y\right)\sigma_{ia}(t)+\frac{1}{2}\ln \coth\left(\frac{\beta\Gamma}{M}\right)\sigma_{ia}(t)\sigma_{ia}(t+1)\right\}\nonumber\\
&=\prod_{i=1}^N\sum_{w_{i}=\pm1}\frac{1}{2}\int Dz \left(\int Dy2\cosh\sqrt{\left(w_i\tilde{m}+\sqrt{\tilde{q}}z+\sqrt{\tilde{R}-\tilde{q}}y\right)^2+(\beta \Gamma)^2}\right)^n\nonumber\\
&\simeq \exp\left\{nN\int Dz \ln 2Y\right\}.
\label{A16}
\end{align}
Here, we introduce the Hubbard-Stratonovich transformation 
\begin{align}
\exp\left(\frac{x^2}{2}\right)=\int Dz\exp\left(xz\right),
\label{A17}
\end{align}
to the terms $(\sqrt{\tilde{q}}/M\sum_{a,t}\sigma_{ia}(t))^2/2$ and $\sum_{a}(\sqrt{\tilde{R}-\tilde{q}}/M\sum_{t}\sigma_{ia}(t))^2/2$.
We apply the inverse operation of the Suzuki-Trotter decomposition and take the trace.

Under the RS ansatz and static approximation, $e^{G_3}$ is represented as
\begin{align}
e^{G_3}&=\exp\left\{nN\left(-m\tilde{m}-\frac{1}{2}R\tilde{R}-\frac{n-1}{2}q\tilde{q}+\mathcal{O}\left(\frac{1}{M}\right)\right)\right\}.\label{A18}
\end{align}
In the thermodynamic limit $N\rightarrow \infty$, the saddle-point method can be applied and the RS free energy density can be expressed as 
\begin{align}
-\beta f_{\mathrm{RS}}&=\underset{\substack{m,q,R\\ \tilde{m},\tilde{q},\tilde{R}}}{\mathrm{extr}}\left[2\alpha \int DuH\left(-X_1\right) \ln \int D\nu \exp\left(-\beta H\left(X_2\right)\right)+\int Dz\ln 2Y-m\tilde{m}-\frac{1}{2}R\tilde{R}+\frac{1}{2}q\tilde{q}\right],\label{A19}
\end{align}
where the order parameters and their auxiliary parameters can be determined by the saddle-point condition in the RS free energy density.

At the classical limit $\Gamma \rightarrow 0$ and  $R\rightarrow 1$, we can obtain the classical RS free energy density.
Depending on the sign of $X_2$, $\exp\left(-\beta H\left(X_2\right)\right)$ term in Eq.\eqref{A15} takes $1$ or $e^{-\beta}$.
This relationship yields
\begin{align}
e^{G_1}&= \exp\left\{2\alpha nN\int DuH\left(-\sqrt{\frac{m^2}{q-m^2}u}\right)\ln \int D\nu\left\{\exp\left(-\beta\right)+\Theta\left(\sqrt{q}u+\sqrt{1-q}\nu\right)\left(1-\exp\left(-\beta\right)\right)\right\}\right\}\nonumber\\
&= \exp\left\{2\alpha nN\int DzH\left(\sqrt{\frac{m^2}{q-m^2}u}\right)\ln \left\{\exp\left(-\beta\right)+\left(1-\exp\left(-\beta\right)\right)H\left(\sqrt{\frac{q}{1-q}}u\right)\right\}\right\}\label{A20}.
\end{align}

We compute $e^{G_2}$ as 
\begin{align}
e^{G_2}&= \exp\left\{nN\left(\int Dz \ln \int Dy 2\cosh\left(\tilde{m}+\sqrt{\tilde{q}}z+\sqrt{\tilde{R}-\tilde{q}}y\right)\right)\right\}\nonumber\\
&=\exp\left\{nN\left(\frac{\tilde{R}-\tilde{q}}{2}+\int Dz\ln 2\cosh\left(\tilde{m}+\sqrt{\tilde{q}}z\right) \right)\right\},\label{A21}
\end{align}
where we utilise the relationship $\int Dx \cosh(ax+b)=e^{a^2/2}\cosh b$.
At the classical limit, $e^{G_3}$ is expressed as
\begin{align}
e^{G_3}&\simeq \exp\left\{nN\left(-m\tilde{m}-\frac{1}{2}\tilde{R}-\frac{n-1}{2}q\tilde{q}\right)\right\}. 
\label{A22}
\end{align}
The classical RS free energy density can be recovered as 
\begin{align}
-\beta f_{\mathrm{RS}}&=\underset{\substack{m,q\\ \tilde{m},\tilde{q}}}{\mathrm{extr}}\left[2\alpha \int DzH\left(\sqrt{\frac{m^2}{q-m^2}u}\right)\ln \left\{\exp\left(-\beta\right)+\left(1-\exp\left(-\beta\right)\right)H\left(\sqrt{\frac{q}{1-q}}u\right)\right\}\right.\nonumber\\
  &\left.+\int Dz\ln 2\cosh\left(\tilde{m}+\sqrt{\tilde{q}}z\right)-m\tilde{m}-\frac{q(1-\tilde{q})}{2}\right].
  \label{A23}
\end{align}
This result is consistent with the results in Ref.\cite{Seung_1992}. 
\end{widetext}
\bibliography{main.bib}
\end{document}